\documentclass[useAMS,usenatbib]{mn2e}
\usepackage{graphicx}
\usepackage{rotating}
\usepackage{pdflscape}

\title[Hydrogen-loss rates of protoplanetary cores]
  {EUV-driven mass loss of protoplanetary cores with hydrogen-dominated atmospheres:
  The influences of ionization and orbital distance}
\author[N. V. Erkaev  et al.]
{N.~V.~Erkaev,$^{1,2}$ H.~Lammer,$^3$ P.~Odert,$^{3}$ K.~G.~Kislyakova,$^{3}$
\newauthor  C.~P. Johnstone,$^4$ M.~G\"{u}del,$^4$ M.~L.~Khodachenko$^3$
 \\
 $^1$Institute of Computational Modelling SB RAS, 660036, Krasnoyarsk, Russian Federation\\
$^2$Siberian Federal University, Krasnoyarsk, Russian Federation\\
$^3$Space Research Institute, Austrian Academy of Sciences, Schmiedlstr. 6, A-8042, Graz, Austria\\
 $^4$Institute for Astronomy, University of Vienna, T\"{u}rkenschanzstrasse 17, 1180 Vienna, Austria}
\date{Released 2015}

\pagerange{\pageref{firstpage}--\pageref{lastpage}} \pubyear{2015}

\def\LaTeX{L\kern-.36em\raise.3ex\hbox{a}\kern-.15em
    T\kern-.1667em\lower.7ex\hbox{E}\kern-.125emX}

\begin{document}

\label{firstpage}

\maketitle

\begin{abstract}
We investigate the loss rates of the hydrogen atmospheres of terrestrial
planets with a range of masses and orbital distances by assuming a stellar extreme ultraviolet (EUV)
luminosity that is 100 times stronger than that of the current Sun.
We apply a 1D upper atmosphere radiation
absorption and hydrodynamic escape model that takes into account ionization, dissociation
and recombination to calculate hydrogen mass loss rates. We study the effects of the ionization,
dissociation and recombination on the thermal mass loss rates of hydrogen-dominated super-Earths
and compare the results to those obtained by the energy-limited escape formula which is widely
used for mass loss evolution studies. Our results indicate that the energy-limited formula can
to a great extent over- or underestimate the hydrogen mass loss rates by amounts that depend on
the stellar EUV flux and planetary parameters such as mass, size, effective temperature, and EUV
absorption radius.
\end{abstract}


\section{INTRODUCTION}
During the early stages of planet formation, protoplanetary cores that are still embedded in the circumstellar disk can accumulate hydrogen-dominated primordial envelopes from the gas disk (e.g., Hayashi et al. 1979; Nakazawa et al. 1985; Wuchterl 1993; Ikoma and Genda 2006; Rafikov 2006; St\"{o}kl et al. 2015a; 2015b). The amount of gas captured by the planetary core strongly depends on its mass. Sufficiently massive cores can end up in a runaway accretion regime leading to subsequent formation of gas giants.

A surprise is that a large number of low mass exoplanets discovered to date by ground-based and space-based facilities such as HATNed (Bakos et al. 2004), SuperWASP (Pollacco et al. 2006), CoRoT (Auvergne et al. 2009) and Kepler (Borucki et al. 2010) was the discovery of a large number of hydrogen-dominated sub-Neptune-type planets at very close orbital distances. Before the discovery of these planet's with masses between 1$M_{\rm \oplus}$ and 10$M_{\rm \oplus}$ at orbit locations $<$0.1 AU from their host stars, it was expected that planets with such low masses should not have primordial hydrogen envelopes and resemble large Mercury-type rocky bodies such as the first discovered planet within this size-mass regime CoRoT-7b (L\'{e}ger et al. 2009).

Owen \& Jackson (2012) were the first who also studied the evaporation of hydrogen by stellar extreme ultraviolet (EUV) and X-ray radiation from planets within the hot `super-Earth' and hot Neptune domain by applying hydrodynamical equations at orbit locations $<$0.1 AU. In this work, the authors assumed X-ray luminosities similar to those observed around young solar-like stars (e.g., $L_{\rm X} \approx 10^{30}$ erg s$^{-1}$) and discovered that close-in H$_2$-dominated planets could experience a X-ray and an extreme ultraviolet (EUV) driven evaporation regime. When X-rays drive the hydrogen escape, the flow passes through a sonic surface and a shock may build up before the ionization front where than EUV heating occurs. When EUV drives the hydrogen escape, a subsonic X-ray flow passes through the ionization front and the EUV heated flow then is either supersonic or proceeds to pass through an EUV heated sonic point.

Owen \& Jackson (2012) found also that the upper atmosphere heating by X-rays, which is related to the photo-electron production by the K-shells of metals where the presence of O and C are important is relevant at orbit locations that are $<$0.1 AU. At orbits that are $>$0.1 AU, heating by EUV photons is the dominant driver for hydrogen escape. From the results of their study, one can see that the transition from X-ray driven to EUV-driven hydrogen escape occurs at lower X-ray luminosities for planets closer to their host stars and for planets with lower densities. The hydrogen mass loss rate for a typical sub-Neptune with
$\approx 1.6 M_{\rm \oplus}$ and $\approx 5 M_{\rm \oplus}$ at 0.1 AU around a young solar like star was in Owen \& Jackson (2012) $\approx 3\times 10^{10}$ g s$^{-1}$. Such a mass loss rate during the early stage
of a solar-like star is similar to those derived by several other groups
(Yelle 2004; Garcia-Mu\~{n}oz 2007; Penz et al. 2008; Murry-Clay et al. 2009; Koskinen et al. 2013a; 2013b; Shaikhislamov et al. 2014; Khodachenko et al. 2015) of $\approx 4-7\times 10^{10}$ g s$^{-1}$ for the hot Jupiter HD~209458b. From these escape models one can conclude that most close-in exoplanets start to evaporate within the X-ray regime but switch to the EUV-driven regime when the X-ray flux falls below a critical value. The X-ray flux is related to the age-activity relation of the planet's host star and the orbital location. From this pioneering study one can also conclude that thermal evaporation is more important for lower mass planets, especially for those who are in the hot Neptune and sub-Neptune domains.

Depending on nebula conditions and the formation scenarios of low mass hydrogen dominated planets, some close-in hot sub-Neptunes and Neptunes may get rid of their initially captured hydrogen envelopes, but similar planets may have a problem to lose them at orbit locations that are $>$0.1 AU. Because such remnants of nebula gas envelopes around super-Earths would be a problem for habitability,
Lammer et al. (2014) investigated the origin and loss of captured hydrogen envelopes from protoplanetary cores with masses in the range of 0.1 to 5.0$M_{\rm \oplus}$ orbiting in the habitable zone at 1 AU of a Sun-like G star. In this study, the authors also applied a 1D hydrodynamic upper atmosphere model and concluded that depending on nebula properties, protoplanetary cores with masses $\le$1 $M_{\rm \oplus}$ orbiting within the habitable zone of a solar-like star most likely cannot lose their captured hydrogen envelopes. Their results have been recently confirmed by Luger et al. (2015) and Owen \& Mohanty (2016) who studied the possibility of a transformation of sub-Neptunes into super-Earths in the habitable zones of M dwarfs determined by the loss of atmospheric hydrogen and Johnstone et al. (2015) who investigated the mass loss of hydrogen envelopes around similar cores as assumed by Lammer et al. (2014) along various stellar rotation related activity evolution tracks in the habitable zones of solar-like stars. From the recent study of Owen \& Mohanty (2016) one finds in agreement with Tian et al. (2005), Erkaev et al. (2013) and Chadney et al. (2015) that the lost envelope mass could be significant lower if one neglects the transition to Jeans escape. Moreover, Owen \& Mohanty (2016) found that cores with masses that are $\ge 1M_{\rm \oplus}$ with initial H$_2$/He envelope mass fractions $\ge 1$\% will not loose their gaseous envelopes during their lifetimes. This finding is also in agreement for similar planets in G-star habitable zones (Lammer et al. 2014).

The results of these studies indicate that depending on the initial nebular properties, such as the dust grain depletion factor, planetesimal accretion rates, and resulting luminosities, protoplanetary cores with masses $\ge$1.0$M_{\rm \oplus}$ orbiting inside the habitable zones of M, K, G and F-type dwarf stars most likely remain or evolve to sub-Neptunes instead of Earth-like planets and keep large fractions of their hydrogen envelopes during their whole life times. As pointed out above, in the meantime, these theoretical results seem to be confirmed by detailed analyses of observations (Marcy et al. 2014; Rogers 2015).

Rogers (2015) analyzed many planets discovered by the Kepler satellite with both radius and mass measurements and concluded that most `super-Earths' with radii of 1.6$R_{\rm \oplus}$ have densities that are too low to be composed of silicates and iron alone. The majority of these low density sub-Neptunes are discovered at closer orbital distances than 1 AU. Taking this into account, in this work we study the hydrogen loss rates from captured gas envelopes with core masses of 1$M_{\rm \oplus}$, 2$M_{\rm \oplus}$, 3$M_{\rm \oplus}$ and 5$M_{\rm \oplus}$ orbiting a moderate rotating young G-star which is 100 times more active in extreme ultraviolet (EUV) radiation compared to the present Sun at orbital distances between 0.1 - 1.0 AU.

The main aim of this study is to investigate how ionization, dissociation, recombination, and Ly-$\alpha$ cooling influences the hydrogen mass loss rates of sub-Neptunes depending on the orbital location between 0.1 and 1 AU around young solar-like stars. Furthermore, we study how the results obtained by the upper atmosphere EUV absorption hydrodynamic escape model differ from those provided by the widely used energy-limited formula (e.g., Lammer et al. 2009; Ehrenreich and D\'{e}sert 2011; Sanz-Forcada et al. 2011; Leitzinger et al. 2011; Lopez et al. 2012; Lopez and Fortney 2013; Valencia et al. 2013; Kurokawa and Kaltenegger 2013; Luger et al. 2015). In Section 2, we describe the modeling approach while the detailed code description is given in an appendix. In Section 3, we discuss the results and summarize our conclusions in Section 4.

\section{MODELING APPROACH}
To study the EUV-heated upper atmosphere structure and thermal escape rates of the hydrogen atoms, we apply an
EUV energy absorption and 1-D upper atmosphere hydrodynamic model applied before in
a more simpler way in several studies (Erkaev et al. 2013; 2014; 2015; Lammer et al. 2013; 2014), and described
in detail in the appendix.
The model solves the system of the hydrodynamic equations for mass, momentum, and energy conservation.
In addition to the mechanisms included in our previous models, the simulations in this study also include
the effects of ionization, dissociation, recombination and Ly-$\alpha$ cooling.

In the present parameter study, which focusses on the mass loss rates, we use an integrated EUV flux and do not consider a wavelength dependence of the incoming stellar EUV radiation. This approach is justified by a recent study of Guo and Ben-Jaffel (2016), who studied the influence of the EUV spectral energy distribution on the upper atmosphere structure, composition and atmospheric escape
of HD~189733b, HD~209458b, GJ~436b and Kepler-11b. These authors applied an EUV spectral energy distribution in their model and found that the total hydrogen mass loss rates are only moderately
affected by the spectral dependence of the EUV flux (see Fig. 13 in Guo \& Ben-Jaffel 2016). From their study one can see that if one considers the hydrogen mass loss rates of the sub-Neptune
Kepler-11b, the variation of the total mass loss rate with the variations of the spectral index remains within a factor 1.33 (Guo \& Ben-Jaffel 2016).
Although no large changes in the total mass loss rates are expected, for understanding the effect of the spectral dependence on the volume heating rate and distribution of different species,
we plan also to apply EUV spectra in the model for future applications. In the present hydrogen mass loss calculations
we assume the EUV luminosity of a moderate rotating young solar-like star (Tu et al. 2015; Johnstone et al. 2015) that is enhanced by a factor of 100 compared to the Sun's
present value with a flux at Earth's orbit of $\approx 4.64$~erg~cm$^{-2}$~s$^{-1}$ (Ribas et al. 2005.

One should note that in H$_2$-dominated atmospheres, the main molecular infra-red (IR) emitting coolant is H$_3^+$.
As shown by Shaikhislamov et al. (2014) and Chadney et al. (2015), this efficient IR-cooling mechanism vanishes or is negligible in hydrogen-dominated upper atmospheres at small orbital distances or high EUV flux values, because due to molecular dissociation preventing the balancing of the stellar heating by IR cooling. Since the assumed EUV flux values related to the moderate rotating young solar-like star after its arrival at the Zero-Age-Main-Sequence (ZAMS) is assumed to be 100 times higher compared to that of today's solar value the flux is high enough, even at 1 AU, that efficient H$_3^+$ IR cooling plays a negligible role for the time period of the studied test planets. After $\approx$100 Myr, when the activity decreases H$_3^+$-cooling will play an important role beyond orbital locations
$> 0.1$ AU (Chadney et al. 2015) and the losses will switch from the hydrodynamical regime to a fast Jeans and Jeans-type loss (Tian et al. 2005; Erkaev et al. 2013; Owen \& Mohanty 2016). These loss rates can be orders of magnitude lower compared to the hydrodynamical-driven loss rates in the early period of the planets life time.

The lower boundary of our simulation domain, \mbox{$R_{\rm 0}=R_{\rm c}+z_0$}, is chosen in a similar way as in Lammer et al. (2014) where $R_{\rm c}$ is the core radius and $z_0$ the altitude of the gas envelope up to the homopause level that is located in the lower part of the thermosphere. From formation and structure models of hydrogen-dominated low mass exoplanets it is known that the optical radius which is caused mainly by H$_2$ Rayleigh scattering and H$_2$-H$_2$ collisional absorption could lie hundreds or thousands of kilometers or even a few Earth-radii above the core radius (e.g., Rogers et al. 2011; Mordasini et al. 2012). The homopause level $z_0$ has lower pressures and lies therefore above the optical radius. However, the homopause level lies at the base of the thermosphere where above this level the bulk of the EUV photons is absorbed and very little penetrates below it. $R_{\rm 0}$ can therefore be considered as a natural boundary between the troposphere-stratosphere-mesosphere and the thermosphere-exosphere.

We assume a hydrogen molecule number density at the lower boundary $R_{\rm 0}$ of $5\times10^{12}$ cm$^{-3}$
(e.g. Kasting \& Pollack 1983; Atreya 1986, 1999; Tian et al. 2005; Erkaev et al. 2013; Lammer et al. 2014). This density value
can never be arbitrarily increased or decreased by as much as an order of magnitude, even if the captured envelope mass fractions
$f_{\rm env}$ of a particular planet are different. The reason for this is that the value of $n_{\rm 0}$ is strictly determined
by the EUV absorption optical depth of the thermosphere.

However, to enable comparison of the hydrogen loss rates at different orbital distances between the test planets we assume fixed lower boundary densities and homopause levels $z_0$ shown in Table 1. The corresponding 1 bar levels (i.e. $R_{\rm 1 bar}$) are also given as a reference.
\begin{table*}
\renewcommand{\baselinestretch}{1}
\caption{Hydrogen mass loss rates for protoplanets with masses of 1$M_{\rm \oplus}$ (a), 2$M_{\rm \oplus}$ (b), 3$M_{\rm \oplus}$ (c) and 5$M_{\rm \oplus}$ (d) with assumed hydrogen envelope mass fractions $f_{\rm env}$ as mentioned in the main text. The hydrogen envelopes are exposed to a stellar EUV flux that is 100 times stronger compared to present solar values at 0.1--1 AU. $L$ is the hydrogen loss rate calculated with the hydrodynamic model neglecting ionization, dissociation and recombination; $L_{n,i}$ is the total escape rate of hydrogen ions and neutrals if ionization, dissociation and recombination are not neglected; $L_{\rm n}$ and $L_{\rm i}$ correspond to the losses of neutral H atoms or H$^+$ ions only. $L_{\rm en}$ and $L_{\rm en}^*$ are the energy limited loss rate cases related to eq. (1) and eq. (3) that have been multiplied by a heating efficiency $\eta$ of 15 \%.}
\begin{center}
\begin{tabular}{cc|cccccccc}
$M_{\rm pl}/M_{\oplus}$ & $R_{\rm 0}/R_{\oplus}$ $|$ $R_{\rm 1 bar}/R_{\oplus}$  & $R_{\rm EUV}/R_{\oplus}$ & $L$ [g~s$^{-1}$] & $L_{\rm n,i}$ [g~s$^{-1}$] & $L_{\rm n}$ [g~s$^{-1}$] & $L_{\rm i}$ [g~s$^{-1}$] & $L_{\rm en}$ [g~s$^{-1}$]& $L_{\rm en}^*$ [g~s$^{-1}$]\\\hline
$d$=1.0 AU & EUV=100 & & &   &  &  &  &   \\
 $T_{\rm eff}$=250 K     &464 erg cm$^{-2}$ s$^{-1}$ &    &    &    &    &     &   & \\ \hline\hline
1 & 1.15 $|$ $\approx 1.0$  & 2.87 &  $2.1\times10^{8}$ & $2.1\times10^{8}$ & $2.0\times10^{8}$ & $8.3\times10^{6}$ & $3.3\times 10^{8}$& $6.8\times 10^{7}$   \\
2 & 2.26 $|$ 1.43 & 5.2 & $8.5\times10^{8}$ & $8.6\times10^{8}$ & $8.2\times10^{8}$ & $5.0\times10^{7}$ & $1.3\times 10^{9}$ & $2.7\times 10^{8}$   \\
3 & 2.44 $|$ 1.72 & 5.12  &$5.8\times10^{8}$ & $5.9\times10^{8}$ & $5.6\times10^{8}$ & $3.0\times10^{7}$ & $9.7\times 10^{8}$ & $2.2\times 10^{8}$   \\
5 & 2.71 $|$ 2.12 & 5.69 & $6.5\times10^{8}$ & $6.7\times10^{8}$ & $6.0\times10^{8}$ & $6.7\times10^{7}$ & $7.2\times 10^{8}$ & $1.8\times 10^{8}$  \\\hline
$d$=0.7 AU & EUV=200 & &  &  &  &  &  & \\
 $T_{\rm eff}$=275 K    &928 erg cm$^{-2}$ s$^{-1}$&    &     &    &    &     &    &\\ \hline\hline
1 & 1.15  $|$ $\approx 1.0$  & 2.64 & $3.6\times10^{8}$ & $3.3\times10^{8}$ & $3.2\times10^{8}$ & $1.3\times10^{7}$ & $7.3\times 10^{8}$ & $1.3\times 10^{8}$  \\
2 & 2.26 $|$ 1.38 & 5.1 & $1.4\times10^{9}$ & $1.2\times10^{9}$ & $1.1\times10^{9}$ & $1.0\times10^{8}$ & $2.7\times 10^{9}$ & $5.2\times 10^{8}$  \\
3 & 2.44 $|$ 1.67 & 5.12  & $8.9\times10^{8}$ & $7.7\times10^{8}$ & $7.0\times10^{8}$ & $6.7\times10^{7}$ & $1.9\times 10^{9}$ & $4.3\times 10^{8}$   \\
5 & 2.71 $|$ 2.07 & 4.87 & $1.1\times10^{9}$ & $1.0\times10^{9}$ & $9.6\times10^{8}$ & $1.0\times10^{8}$ & $1.8\times 10^{9}$ & $3.7\times 10^{8}$   \\\hline
$d$=0.5 AU & EUV=400 &   &  &  &  &   & \\
 $T_{\rm eff}$=325 K    & 1856 erg cm$^{-2}$ s$^{-1}$  &   &     &    &    &     &    &\\ \hline\hline
1 & 1.15 $|$ $\approx 1.0$ & 2.87 & $5.5\times10^{8}$ & $4.8\times10^{8}$ & $4.6\times10^{8}$ & $2.5\times10^{7}$ & $1.7\times 10^{9}$ & $2.8\times 10^{8}$ \\
2 & 2.26 $|$ 1.29 & 5.2 & $2.0\times10^{9}$ & $1.8\times10^{9}$ & $1.6\times10^{9}$ & $1.8\times10^{8}$ & $5.5\times 10^{9}$ & $1.0\times 10^{9}$\\
3 & 2.44 $|$ 1.59 & 5.12  & $1.5\times10^{9}$ & $1.8\times10^{9}$ & $1.3\times10^{9}$ & $1.8\times10^{8}$ & $3.8\times 10^{9}$ & $8.3\times 10^{8}$\\
5 & 2.71 $|$ 2.0 & 5.69 & $2.3\times10^{9}$ & $2.0\times10^{9}$ & $1.7\times10^{9}$ & $3.2\times10^{8}$ & $3.2\times 10^{9}$& $6.7\times 10^{8}$\\\hline
$d$=0.3 AU & EUV=1111 &   &  &  &  &   & \\
 $T_{\rm eff}$=420 K    & 5166 erg cm$^{-2}$ s$^{-1}$   &     &     &    &    &     &    & \\ \hline\hline
1 & 1.15 $|$ $\approx 1.0$ & 2.41 & $2.5\times10^{9}$ & $2.8\times10^{9}$ & $2.2\times10^{9}$ & $6.7\times10^{8}$ & $3.3\times 10^{9}$ &  $7.7\times 10^{8}$ \\
2 & 2.26 $|$ 1.16 & 4.52 & $1.1\times10^{10}$ & $1.2\times10^{10}$ & $7.7\times10^{9}$ & $3.8\times10^{9}$ & $1.2\times 10^{10}$ &  $2.8\times 10^{9}$ \\
3 & 2.44 $|$ 1.45 & 5.36   &  $3.5\times10^{9}$ & $3.1\times10^{9}$ & $2.5\times10^{9}$ & $6.7\times10^{8}$ & $1.2\times 10^{10}$ &  $2.5\times 10^{9}$ \\
5 & 2.71 $|$ 1.85 & 5.96 &    $4.2\times10^{9}$ & $3.5\times10^{9}$ & $2.8\times10^{8}$ &  $6.7\times10^{8}$ & $9.7\times 10^{9}$&  $2.0\times 10^{9}$ \\\hline
$d$=0.1 AU & EUV=10000 &   &    &  &  &   & \\
 $T_{\rm eff}$=730 K    & 46500 erg cm$^{-2}$ s$^{-1}$&    &      &    &    &     &    &\\ \hline\hline
1 & 1.15 $|$ $\approx 1.0$  & 2.41 & $1.5\times10^{10}$ & $1.8\times10^{10}$ & $9.9\times10^{9}$ & $8.2\times10^{9}$ & $3.0\times 10^{10}$ & $6.7\times 10^{9}$ \\
2 & 2.26 $|$ $\approx 1.0$& 3.84 &   $5.7\times10^{10}$ & $7.7\times10^{10}$ & $3.9\times10^{10}$ & $3.9\times10^{10}$ & $7.5\times 10^{10}$ & $2.7\times 10^{10}$ \\
3 & 2.44 $|$ 1.141 & 4.63  & $2.5\times10^{10}$ & $3.5\times10^{10}$ & $1.5\times10^{10}$ & $2.0\times10^{10}$ & $8.0\times 10^{10}$ & $2.2\times 10^{10}$ \\
5 & 2.71 $|$ 1.54 & 5.15 & $1.0\times10^{10}$ & $1.7\times10^{10}$ & $7.3\times10^{9}$ & $1.0\times10^{10}$ & $6.5\times 10^{10}$ & $1.8\times 10^{10}$ \\\hline
\end{tabular}
\end{center}
\end{table*}
In the present study, we also compare the mass loss rates obtained by the above described upper atmosphere model with the widely used energy-limited escape formula
\begin{equation}
L_{\rm en}=\frac{\pi \eta R_{\rm 0}R_{\rm EUV}^2 I_{\rm EUV}}{GM_{\rm pl}},
\end{equation}
where $I_{\rm EUV}$ is the stellar EUV flux outside the atmosphere at the orbital location of the planet, $\eta$ is the heating efficiency, $R_{\rm EUV}$ is the effective radius corresponding to the absorption of the stellar EUV flux in the upper atmosphere, and $G$ is Newton's gravitational constant.
If one considers pure H$_2$ atmospheres the planetary radius would be caused by Rayleigh scattering and H$_2$-H$_2$ collisional absorption close to the 1 bar level (Brown 2001). However, in real planetary atmospheres clouds and hazes may be present which could also extinct visible light at pressure levels between the 1 bar and the homopause level (e.g., Lopez et al. 2012; Lopez and Fortney 2013).
Note that $I_{\rm EUV}$ does not need to be averaged over the planet's surface because eq. (1),
cast in this form, already accounts for this (cf. Erkaev et al. 2007). $R_{\rm EUV}$ depends on the density distribution and can be determined from the following equation (Erkaev et al. 2014; Erkaev et al. 2015)
\begin{eqnarray}
R_{\rm EUV}=R_0 \left[1+2\int_{1}^\infty{[1-J_{\rm EUV}(r,\pi/2)/I_{\rm EUV}] r dr}\right]^{0.5},
\end{eqnarray}
where $J_{\rm EUV} (r,\theta) $ is the stellar EUV flux in the atmosphere as function of the dimensional radius $r=R/R_0$ and spherical angle.
As shown by Watson et al. (1981), $R_{\rm EUV}$ can exceed the planetary radius quite substantially, especially for hydrogen-dominated low mass bodies with low gravity fields and high EUV fluxes. For gas giants and other massive and compact planets, $R_{\rm EUV}$ is close to $R_{\rm 0}$.
Therefore, $R_{\rm EUV}$ is often approximated with $R_{\rm pl}\approx R_0$ in the literature (e.g., Ehrenreich and D\'{e}sert 2011; Luger et al. 2015)
\begin{equation}
 L_{\rm en}^*\approx\frac{\pi \eta R_{0}^3 I_{\rm EUV}}{GM_{\rm pl}}.
\end{equation}
To see the difference between both approaches, we compare the results of both assumptions with the hydrodynamic model results. We use the model described in the appendix and locate hydrogen-dominated protoplanets with masses of 1$M_{\rm \oplus}$, 2$M_{\rm \oplus}$ , 3$M_{\rm \oplus}$  and 5$M_{\rm \oplus}$ and at orbital locations of 0.1 AU, 0.3 AU, 0.5 AU, 0.7 AU and 1 AU and expose the hydrogen envelopes to EUV flux values scaled corresponding to the orbital locations with the EUV luminosity of a moderate rotating young solar-like young star (Tu et al. 2015) that is enhanced by a factor of 100 compared to today's solar value.
\section{RESULTS AND DISCUSSIONS}
\begin{figure*}
\begin{center}
\includegraphics[width=0.95\columnwidth]{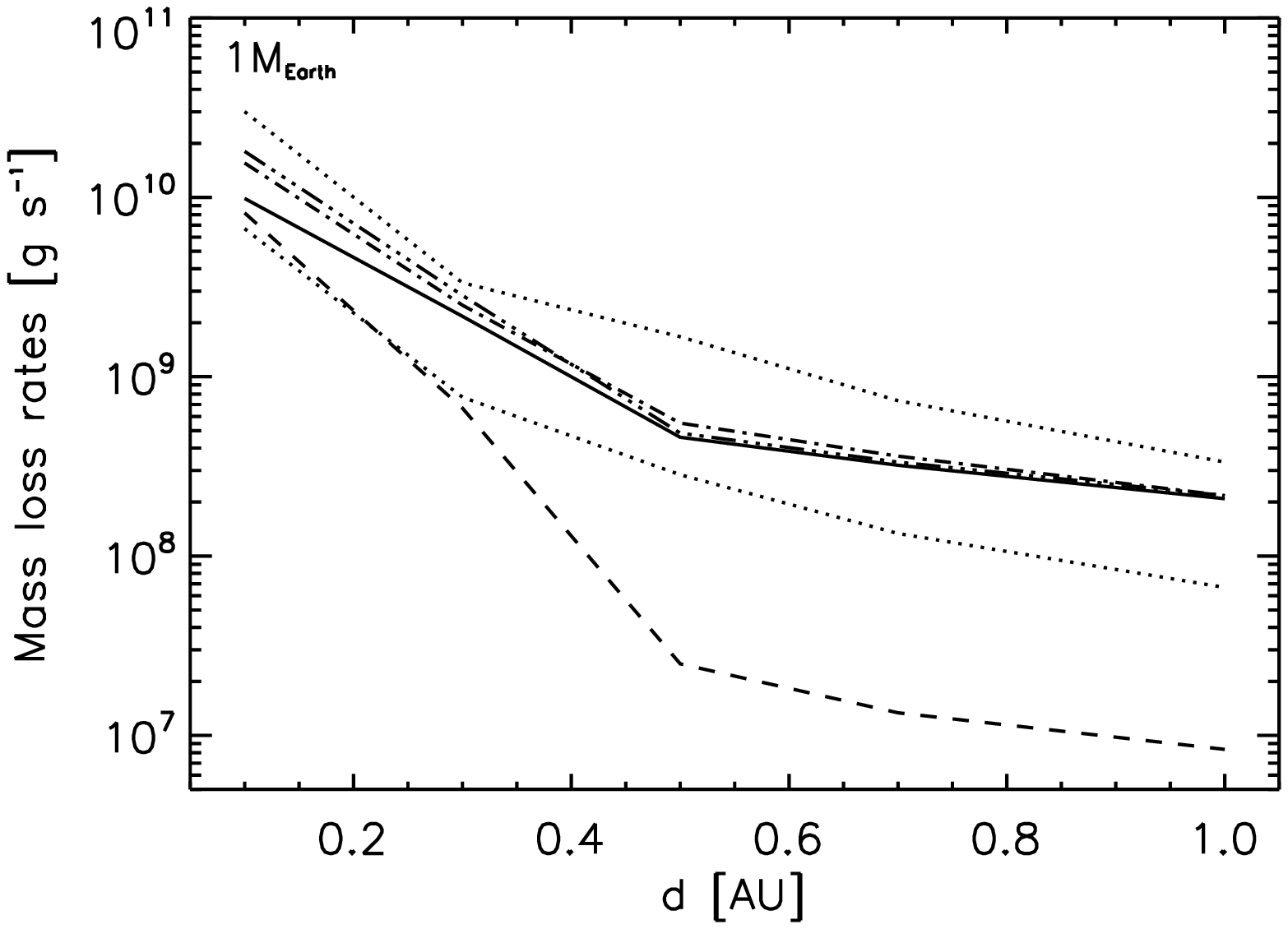}
\includegraphics[width=0.95\columnwidth]{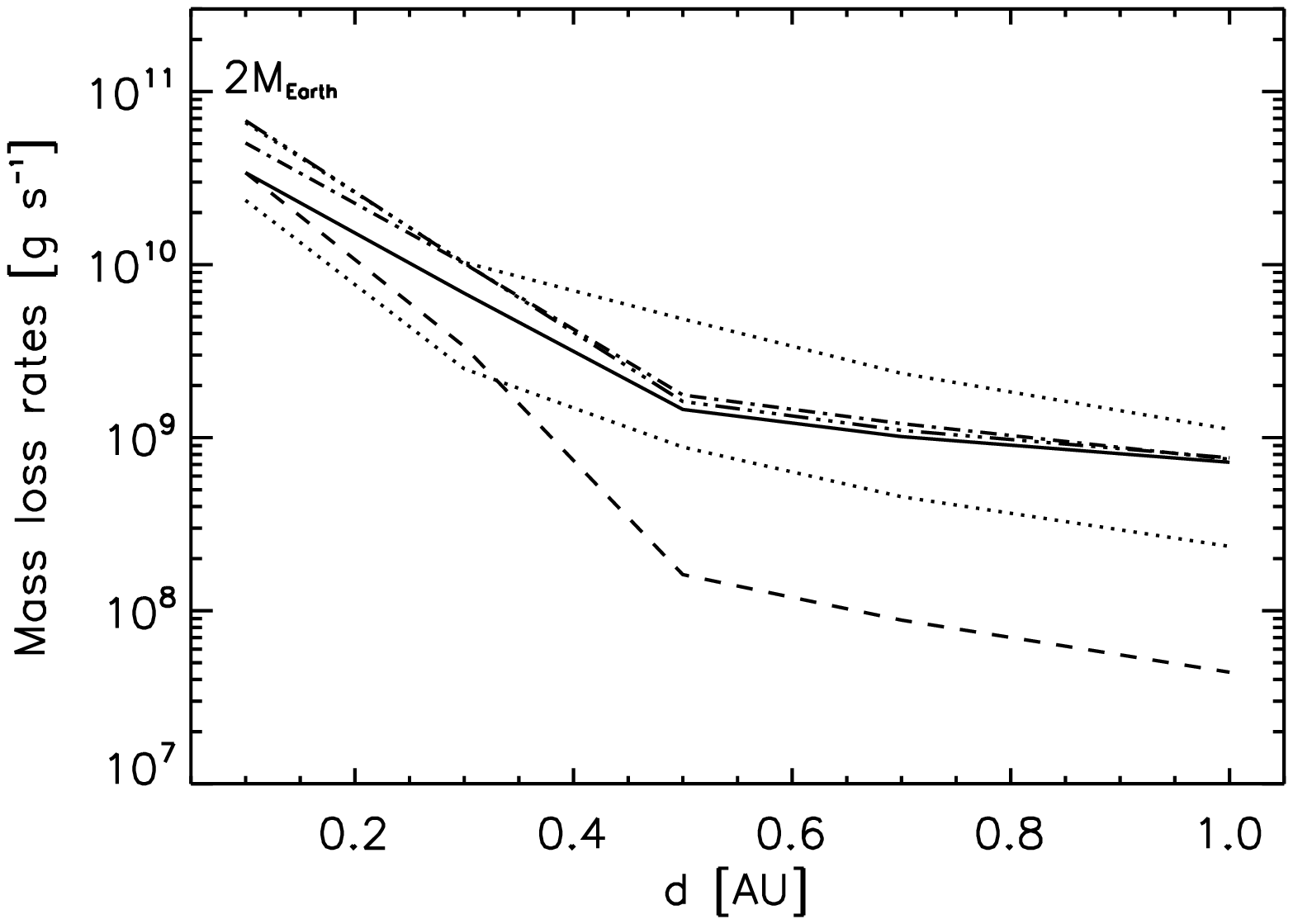}
\includegraphics[width=0.95\columnwidth]{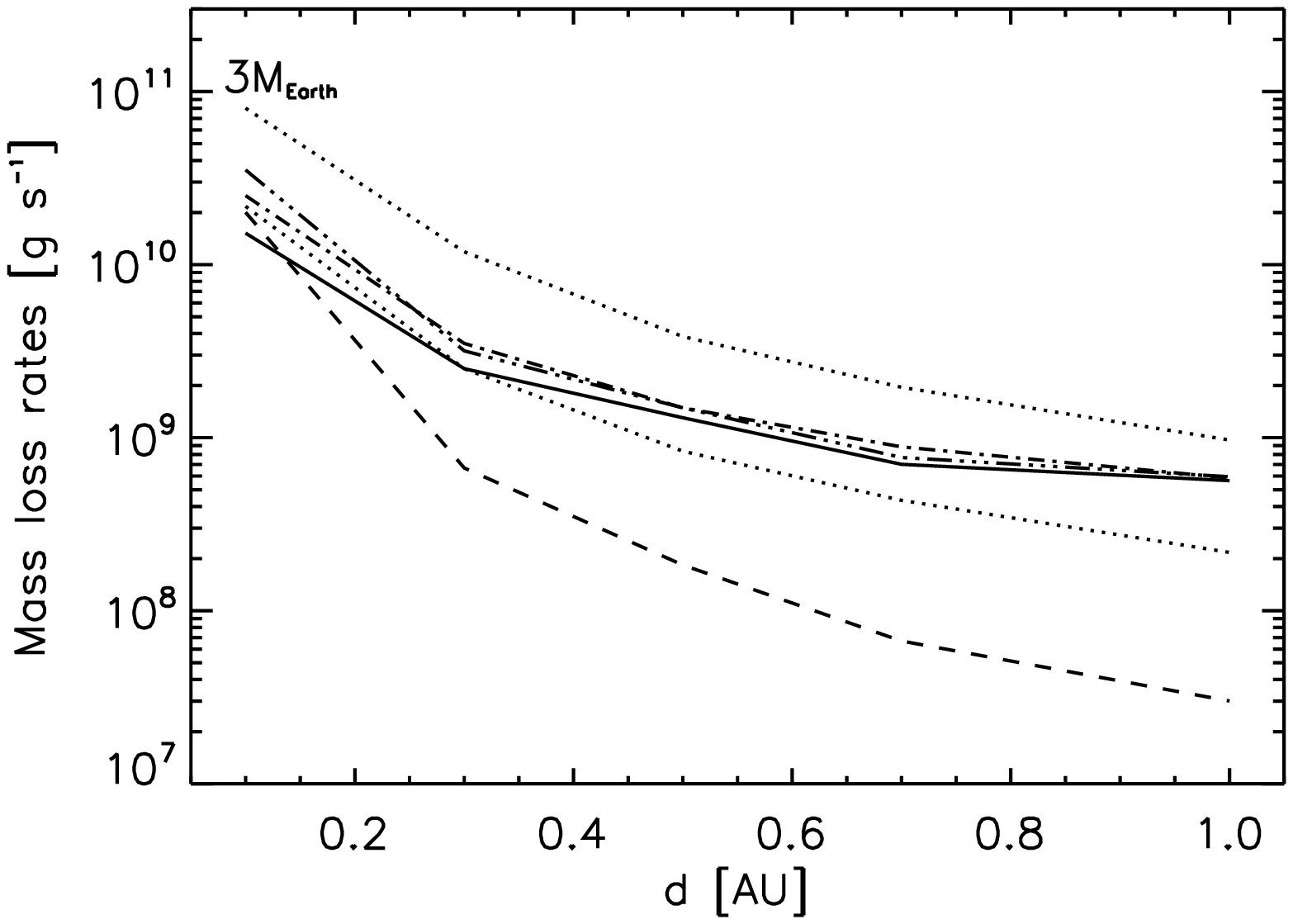}
\includegraphics[width=0.95\columnwidth]{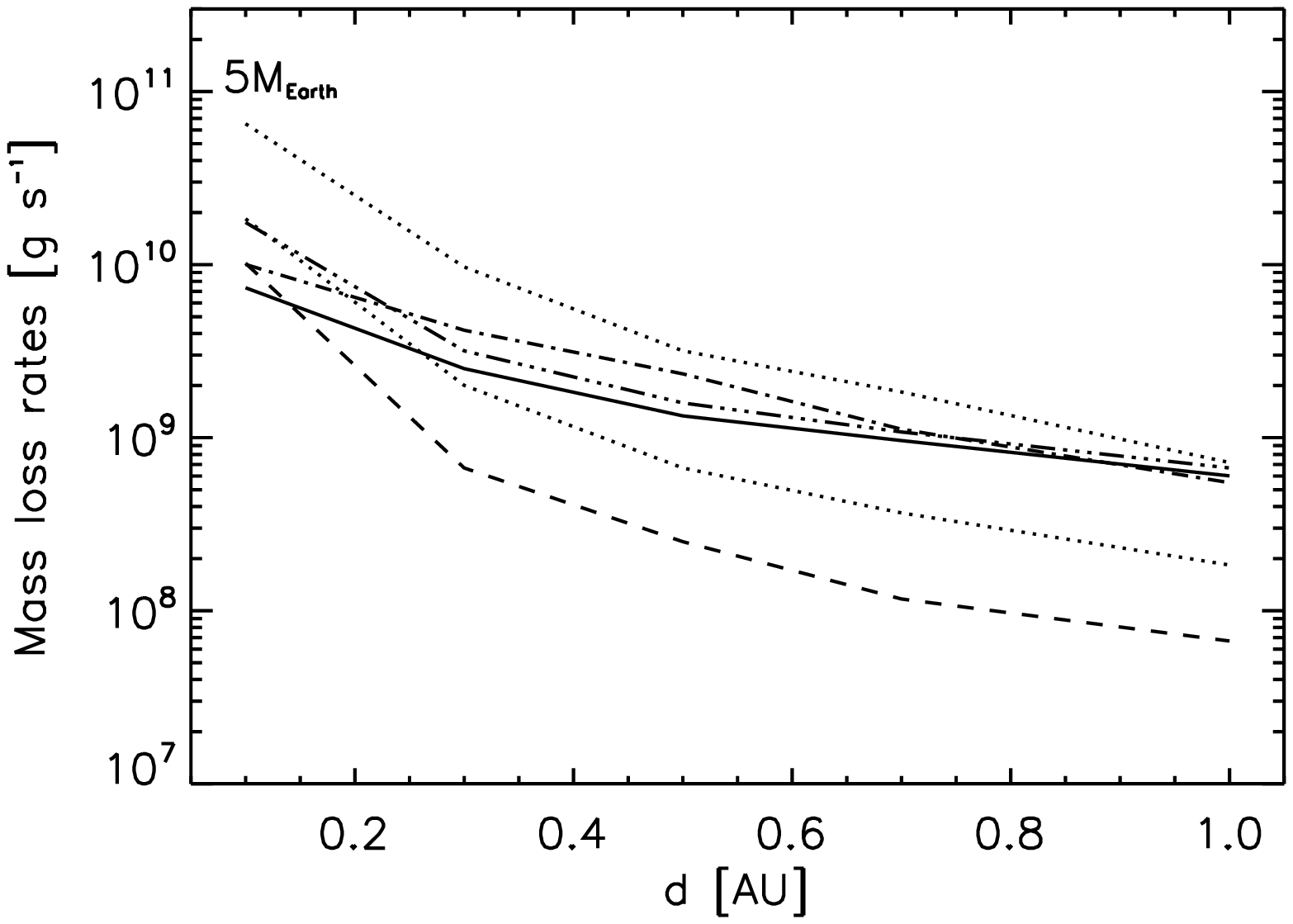}
\caption{Hydrogen mass loss rates as a function of orbital distance for rocky protoplanetary cores with 1$M_{\oplus}$ (a), 2$M_{\oplus}$ (b), 3$M_{\oplus}$ (c) and 5$M_{\oplus}$ (d) calculated for the stellar EUV flux 100 times higher compared to the present solar value in various orbit locations between 0.1 - 1 AU. The planetary radii related to assumed hydrogen envelope fractions are given in Table 1. Dash-dotted lines correspond to the hydrogen loss rate  $L$ of the hydrodynamic model by neglecting ionization, dissociation and recombination; dash-dotted-dotted-dotted lines correspond to loss rates $L_{\rm n,i}$ if ionization, dissociation and recombination
processes are not neglected; dashed lines correspond to the loss rates $L_{\rm i}$ of ionized hydrogen atoms only; solid lines correspond to the loss rates $L_{\rm n}$ of neutral hydrogen atoms only; the upper dotted lines correspond to the energy-limited loss formula multiplied by a heating efficiency $\eta$ of 15 \%, $L_{\rm en}$ according to eq. (1) and the lower dotted lines correspond to the loss rates $L_{\rm en}^*$ according to eq. (3).}
\end{center}
\end{figure*}

Table 1 summarizes the thermal hydrogen mass loss rates from the different protoplanets at different orbital locations. It is important to note that the assumed gas envelope masses are negligible compared to the core masses. Depending on the formation scenarios and nebular conditions, similar cores can capture different amount of nebular gas (e.g., Rogers et al. 2011; Mordasini et al. 2012; St\"{o}kl et al. 2015). If the captured envelope mass was larger, then $R_{\rm 0}$ would also move to larger distances. It was shown in Lammer et al. (2014) that in such cases, the mass loss rates would also be higher. As the gas envelope evaporates, $R_{\rm 0}$ shrinks and as a consequence the mass loss rate also decreases. Because of this effect if one models the mass loss over time the shrinking of $R_0$ has to be considered (Johnstone et al. 2015). Furthermore, it was shown by Tu et al. (2015) that depending on the initial rotation rate the EUV activity levels of young solar-like stars can evolve very differently during the first Gyr of their life time. We do not study here the mass loss of the test planets for the whole range of possible EUV evolutionary scenarios. The hydrogen mass loss rates shown in Table~1 represent therefore only a phase during the planet's evolution.
\begin{figure*}
\begin{center}
\includegraphics[width=0.95\columnwidth]{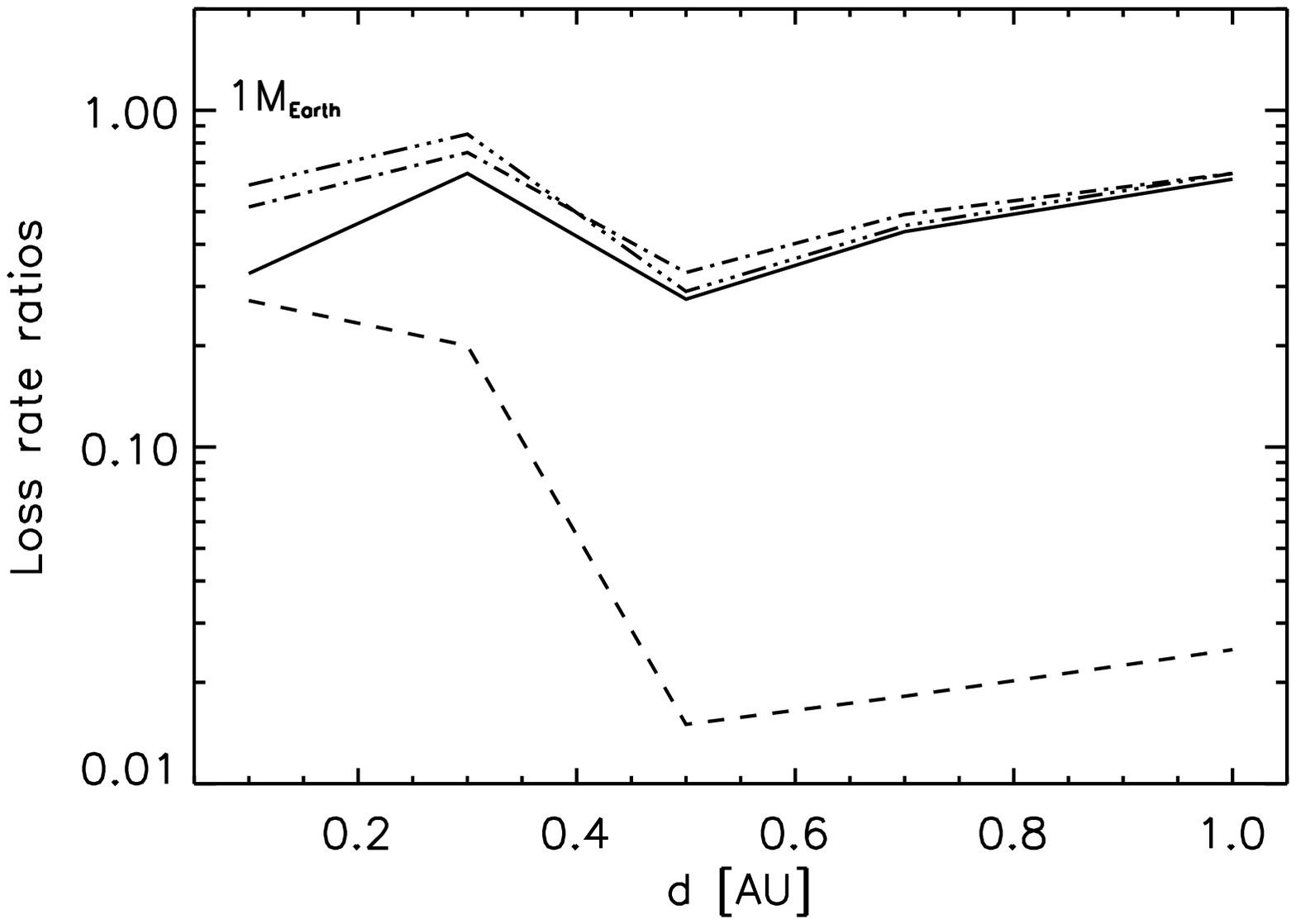}
\includegraphics[width=0.95\columnwidth]{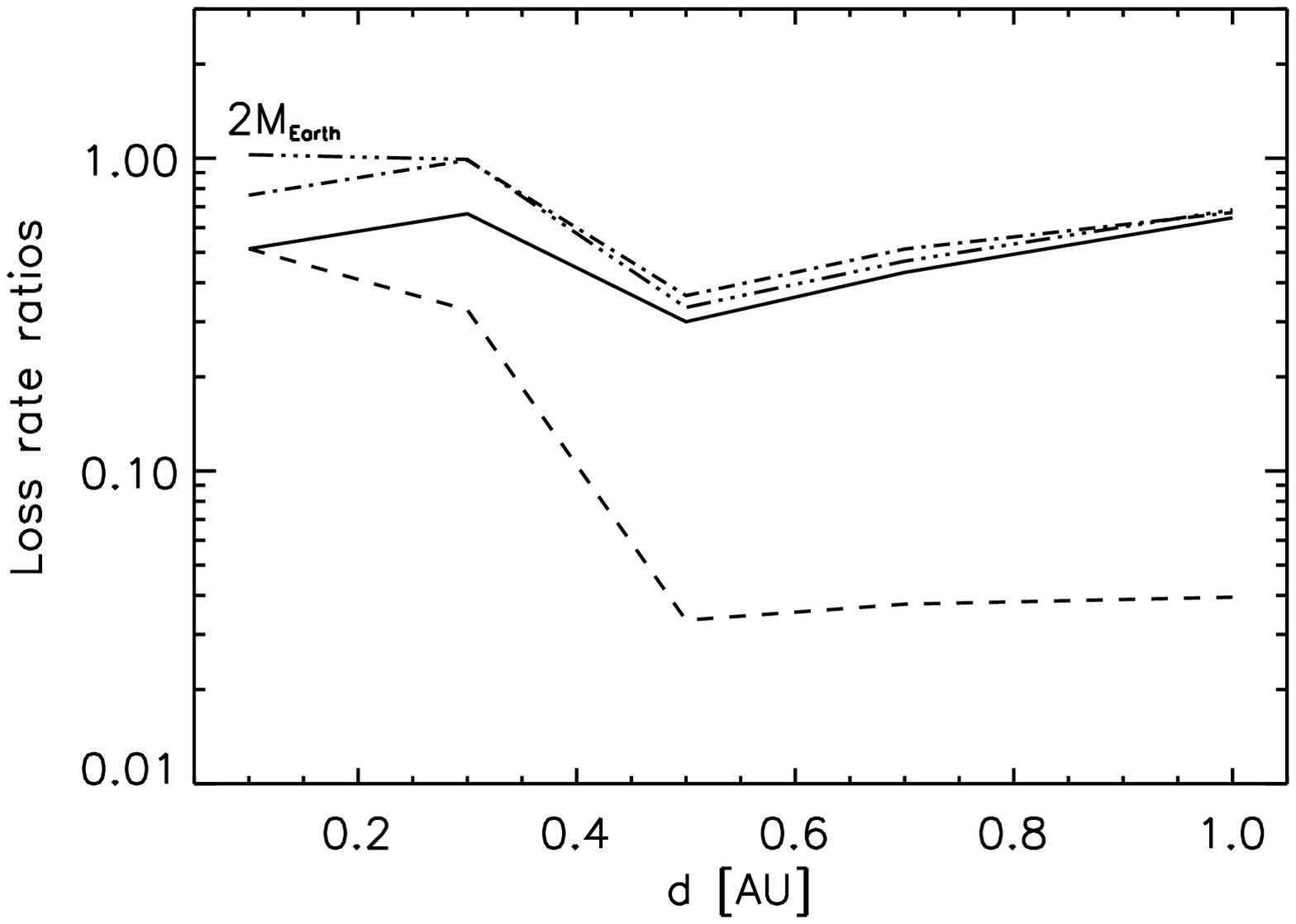}
\includegraphics[width=0.95\columnwidth]{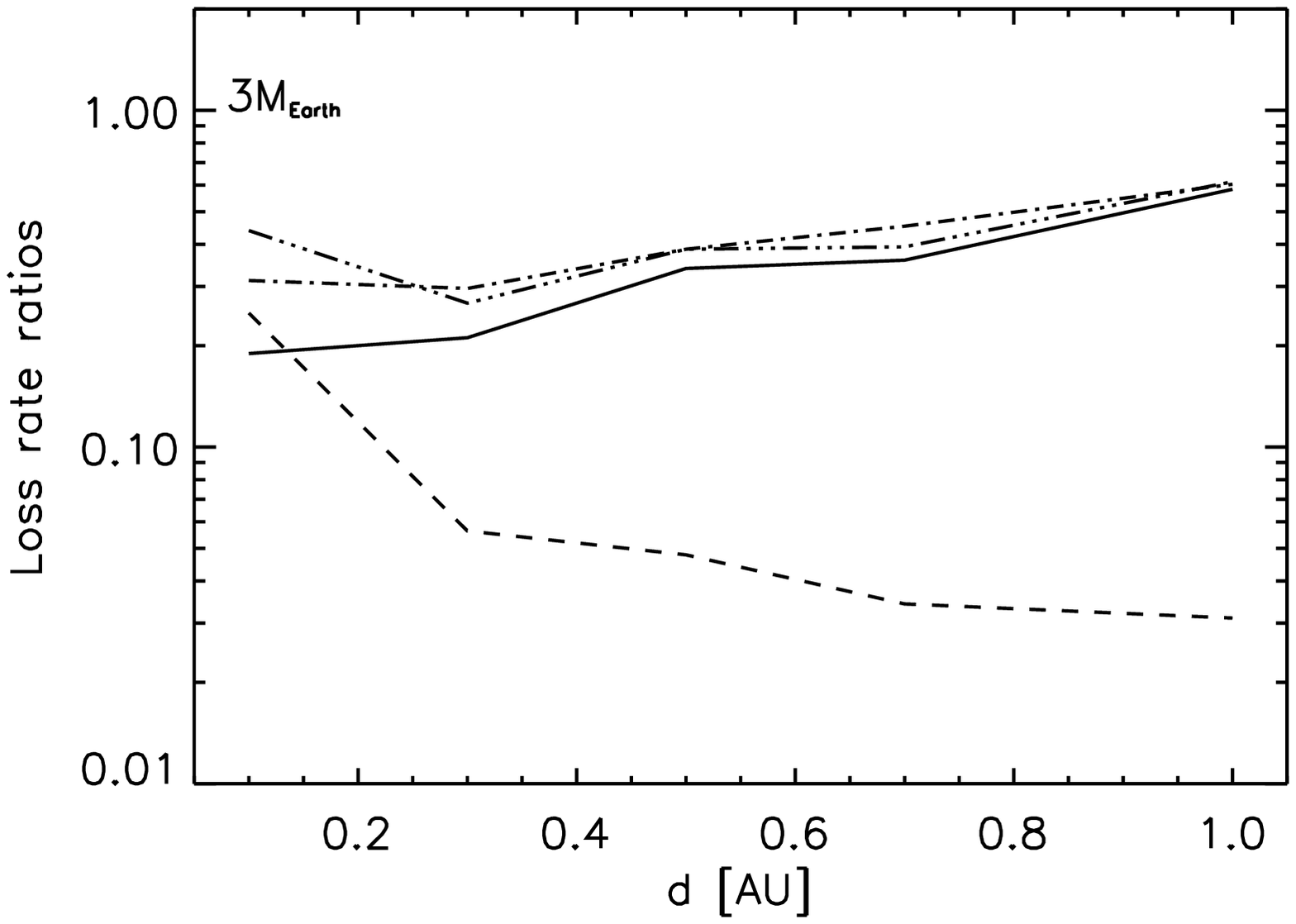}
\includegraphics[width=0.95\columnwidth]{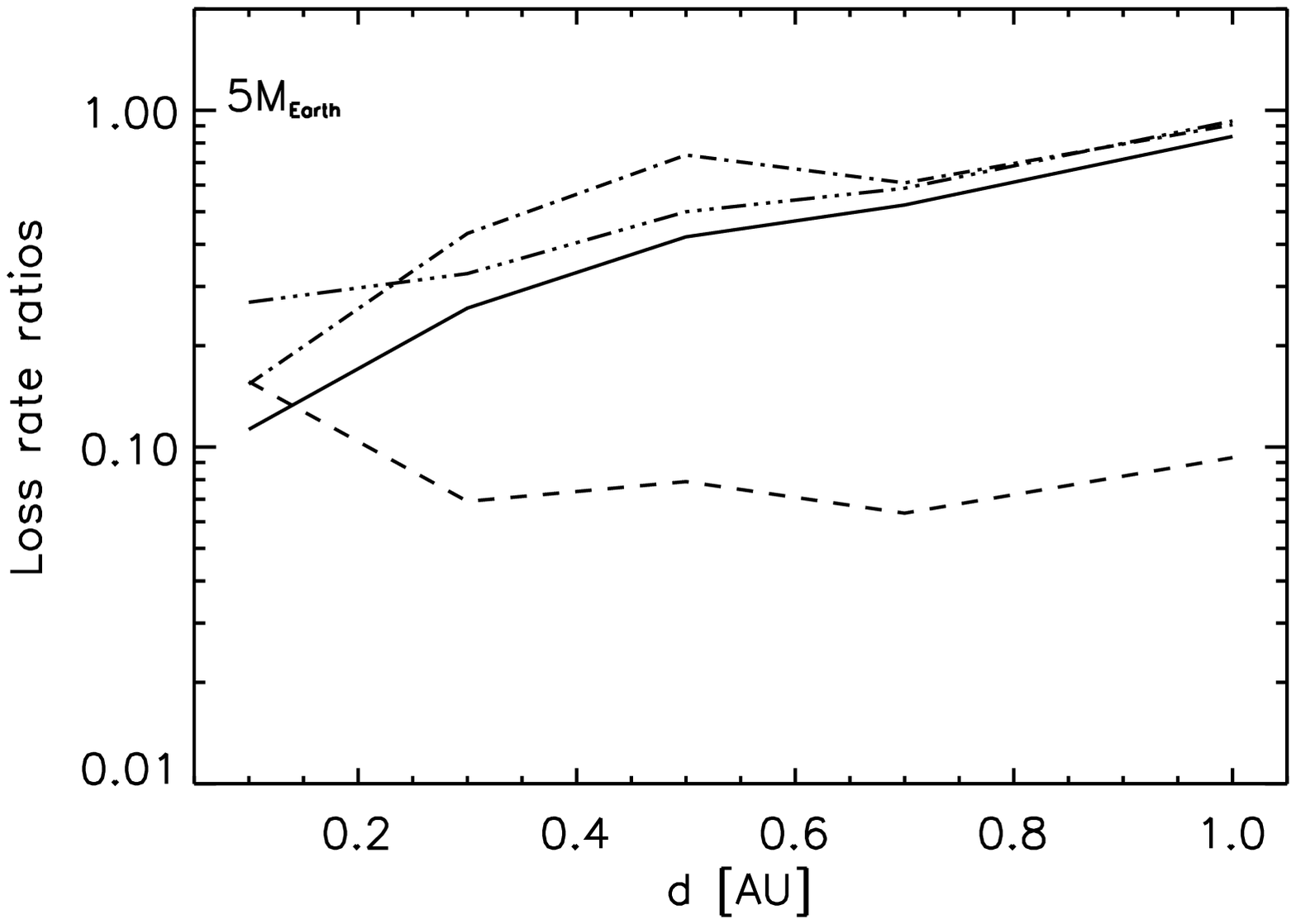}
\caption{Hydrogen mass loss rates, normalized to that obtained by the energy-limited formula $L_{\rm en}$ of eq. (12). Dash-dotted lines: the loss rates $L$ of the hydrodynamic model by neglecting ionization and recombination processes; Dash-dotted-dotted lines: loss rates of $L_{\rm n,i}$; dashed lines: ion loss rates $L_{\rm i}$ only; solid lines: neutral atom loss rates $L_{\rm n}$ only.}
\end{center}
\end{figure*}

Table~1 shows hydrogen mass loss rates calculated by using the described hydrodynamic model for all protoplanets at the studied orbital locations. One can also see that the 1 bar level for a H$_2$ envelope with a homopause distance at 1.15$R_{\oplus}$ and a core mass of 1$M_{\oplus}$ would be located near the core radius if one assume that the atmosphere is isothermal below the homopause level. In such cases, with the corresponding loss rates the thin hydrogen envelope would be lost immediately if no additional source from other volatiles such as H$_2$O and/or CH$_4$ is present. For each orbital location and planetary mass, we run our hydrodynamic simulations twice with different physical mechanisms included. In the first set of simulations, we neglect ionization, dissociation, and recombination, with the mass loss rates being given by $L$. In the second set of simulations, we include the effects of ionization, dissociation, and recombination, with ($L_{n,i}$) giving the total mass loss rates, and $L_{\rm n}$ and $L_{i}$ giving the mass loss rates for neutrals and ions separately.

For comparing the mass loss rates with the energy-limited formula, we show also mass loss rates, $L_{\rm en}$, obtained by this approach, but multiplied with a heating efficiency $\eta$ of 15 \%.
Depending on the planetary parameters, the orbital distance, corresponding EUV flux and effective temperature, the hydrogen mass loss rates are between $\sim 10^{8}$ g s$^{-1}$ at 1 AU and $\sim 10^{10}$ g s$^{-1}$ at 0.1 AU. One can also see that $L_{\rm n,i}$ yields negligible differences that are less than a factor of two for all test planet loss rates between the hydrodynamic upper atmosphere model and the energy-limited formula of eq. (1) at orbital distances of 1 AU. For closer orbits such as 0.5 AU or 0.1 AU, depending on planetary parameters, the differences between the energy-limited formula given in eq. (1) and the results obtained for $L_{\rm n,i}$ are factors of $\approx$1.5--3.5 and $\approx$3.0--9.0, respectively. Therefore, ionization, dissociation and recombination can not be neglected for close orbital distances or highly active young host stars.

Fig. 1 shows the hydrogen mass loss rates for the four planetary masses as a function of orbital distance. One can see that the energy-limited formula underestimates the mass loss rates when one assumes
$R_{\rm EUV}\approx R_{0}\approx R_{\rm pl}$ (eq. 3) and overestimates the mass loss rates when one uses the more accurate formula of eq. (1). As it is obvious also from Figs 1., eq. (3) tends to underestimate the mass loss rates because of the assumption that the EUV flux is absorbed close to the planetary radius, whereas it is actually absorbed at larger altitudes. For the small planets considered here, $R_{\rm EUV}$ may be located at $2-3\,R_0$, much higher than for more massive giant planets (Erkaev et al. 2007; Murray-Clay et al. 2009). Therefore, the application of the energy-limited formulae given in eqs. (1) and (3) are of limited use for low-mass planets.

Moreover, the discrepancy between the mass loss rates calculated with the hydrodynamic code and the energy limited escape formula arises because eq. (1) yields the maximum EUV-driven mass-loss rate that a planet can have, even if multiplied by the heating efficiency $\eta$. The numerator represents the integrated EUV heating rate provided to the planet, i.e. the total stellar EUV flux absorbed at $R_{\rm EUV}$ multiplied by the heating efficiency, i.e., the fraction of incident energy converted to heating. Since the denominator represents the potential energy of the planet, eq. (1) assumes that the total absorbed EUV energy is used to lift the planetary atmosphere out of the planet's gravitational well (Lammer et al. 2016). However, in transonic escape, some fraction of the absorbed energy is also converted to kinetic and thermal energy.
As shown by Johnstone et al. (2015), one of the reasons for this is that
in a transonic wind, a large part of the input energy is absorbed in the
supersonic part of the wind and therefore cannot contribute to the mass
loss rate. In these cases, additional terms increase the denominator and reduce the atmospheric mass loss rate (Sekiya et al. 1980; Johnson et al. 2013; Erkaev et al. 2007; 2015). For certain planetary and stellar parameter combinations, these terms are not negligible leading to a true mass loss rate, as determined with a hydrodynamic model, lower by a factor of a few than those from eq. (1).

Koskinen et al. (2014) therefore suggested to replace $\eta$ with a mass loss efficiency factor to account for these discrepancies. However, it is difficult to estimate this factor since it depends on planetary and stellar parameters, as illustrated by the variation of this discrepancy for different planets and orbits. However, for hot Jupiters such a scaling has been implemented recently by Salz et al. (2016). On the other hand, for increasing $T_{\rm eff}$ the hydrodynamic mass-loss rates increase, which is also not accounted for in eq. (1) (Johnson et al. 2013; Erkaev et al. 2015).
The energy-limited mass loss rates should always be higher than those obtained by the hydrodynamic model.
The only exception is if the planetary equilibrium temperature due to the
star's entire radiation field (i.e. its bolometric luminosity) is so high
that the atmosphere's thermal energy overcomes the gravitational potential
of the planet in regions lower than where EUV heating is taking place (Owen \& Wu 2015; Lammer et al. 2016; Owen \& Subhanjoy 2016).

One can also see that the inclusion of ionization, dissociation and recombination has only a small effect at the assumed orbital distances if one assumes that all neutral atoms and ions can escape from the planets.
Only for the more massive planets and extreme high EUV fluxes at close orbital distance ($< 0.15$~AU), the number density of ions reaches the same value as the neutrals.
Our results indicate also that by including collisional ionization additionally to ionization caused by the stellar EUV flux, the mass loss rates are not affected significantly for orbital locations $\ge$ 0.1 AU. The same can be said for Lyman-$\alpha$ cooling. An inclusion of Lyman-$\alpha$ cooling has only a small effect on the mass loss rates for the closest test planets.
Different mass loss rates related to neutrals and ions depend strongly on the planetary parameters. Ionization becomes more relevant if the planet is massive and, as a consequence, the upper atmosphere is more compact. Ionization also influences the total mass loss rates because a high degree of ionization reduces the area of the neutral atoms where the stellar EUV flux can be absorbed and heat transferred to neutrals.
Figs. 1 shows also a clear break in the slope of the mass loss rates with a separation around 0.5 AU. The reason for this break is a strong nonlinear dependence related to the EUV flux $I_{\rm EUV}$ and the rise of $T_{\rm eff}\approx T_0$. The increase of I$_{\rm EUV}$ and $T_0$ at closer orbital distances lead to a strong increase of the hydrogen loss rates. Because we study only two orbit locations around 0.5 AU, the behavior of the curve looks
like a break of the slope, while in reality the behavior would be smoother. Moreover, one should note that the above mentioned effect depends strongly on the stellar EUV flux and planetary parameters.

The mass loss rates of the four test planets considered at 0.1 AU are comparable to those of hot Jupiters at 0.045~AU (e.g., Yelle 2004; Koskinen et al. 2013; Shaikhislamov et al. 2014; Khodachenko et al. 2015). If we compare for instance the loss rate of our 3$M_{\rm \oplus}$ test planet, we obtain a hydrogen loss rate of $\approx 3.5 \times 10^{10}$ g s$^{-1}$ which is in agreement with the loss rate of a similar planet given in Fig. 5 of Owen \& Jackson (2012). Recent studies by Howe and Burrows (2015) studied mass loss rates from low density exoplanets with a coupled planetary structure and mass loss model which is based on the energy limited formula given in eq (3). These authors exposed sub-Neptune's with similar EUV fluxes as in our study and obtained mass loss rates for hydrogen envelope mass fractions $\le 0.01$ in the order of $\le 10^9$ g s$^{-1}$ at 0.1 AU and $\le 10^8$ g s$^{-1}$ at 0.3 AU. In our hydrodynamic model simulations we obtained for such scenarios mass loss rates which are an order of magnitude higher, but the estimates with eq. (3) yield comparable results.

As mentioned above, our results represent only a finite window of possibilities and would be different if the young star was a slow or fast rotator, meaning lower or higher EUV fluxes than assumed in this study. Furthermore, different accumulated nebula gas masses would also change the mass loss rates. If the planets had magnetic moments and resulting magnetospheres, the high degree of ionization at close orbital distances would also reduce the total mass loss rates (Khodachenko et al. 2015). The discovery of many small close-in low density planets at orbital distances $<0.2$ AU (Marcy et al. 2014; Rogers 2015) indicates that these objects may have evolved from initially more massive planets to sub-Neptunes and hydrogen-dominated super-Earths, but have never lost their envelopes completely. On the other hand, their host stars could also have been less active stars when they were young.

Fig. 2 shows the mass loss rates of neutrals only, ions only and the sums of neutrals and ions, and hydrodynamic loss rates that consider no ionization, dissociation and recombination, normalized to that corresponding to the energy-limited mass loss rate $L_{\rm en}$ (eq. 12). For very high EUV fluxes, ionization alters the mass loss rates because the increasing number of electrons enhances recombination leading to a large fraction of the energy being lost by cooling radiation (Murray-Clay et al. 2009; Guo 2011). In these studies, for Jupiter-type planets this becomes important for EUV fluxes \mbox{$>10^4$ erg cm$^{-2}$ s$^{-1}$}. For considered low mass planets the mass loss rates with ($L_{\rm n,i}$) and without ionization ($L$) are very similar and very small deviations occur only for the closest orbits. This effect would likely become more relevant for even closer orbits or higher stellar EUV emission. In such case, eqs. (12) or (14) are not applicable and approximate estimates for mass loss rates in radiation/recombination-limited regime can be used (Murray-Clay et al. 2009; Owen and Jackson 2012; Luger et al. 2015). One can also see that the rise in ionization occurs for more massive planets with compact upper atmospheres at closer orbital distances, higher effective temperatures and higher EUV fluxes compared to lower mass planets with less compact upper atmospheres.

After having some idea how ionization, dissociation and recombination influence the atmospheric mass loss of hydrogen envelopes around various protoplanetary cores, one can
investigate the orbital locations where `naked' super-Earths or sub-Neptunes which lost their captured nebular gas can be expected. If we use the mass loss rates from Table 1 and estimate roughly how much atmosphere could be lost during the first 100 Myr after the protoatmosphere capture (Lammer et al. 2014) one finds that depending on nebula parameters such as the dust depletion factor $f\approx 0.01$ and assumed relative accretion rates
$\frac{\dot{M}_{\rm acc}}{M_{\rm pl}}$  (yr$^{-1}$) of $\approx 10^{-6}$, cores with masses of $\le 2M_{\rm \oplus}$ can lose their captured envelopes
related to the assumed $f_{\rm env}$, EUV flux most likely within orbital distances that are $\le$ 0.3 AU.
If the accretion rate is $\approx 10^{-7}$, more massive envelopes can be captured, which would then only be lost at orbital locations $\le$ 0.1 AU. A higher dust depletion factor $f\approx 0.1$ in combination with accretion rates that are $< 10^{-6}$ could remove less massive envelopes from a protoplanetary core with $\le 2M_{\rm \oplus}$
even at Venus orbit at 0.7 AU. More massive cores if they originate within the nebula lifetime will keep a fraction of their captured hydrogen envelope even at orbital locations of 0.1 AU.

However, a detailed study taking into account the complete parameter space to determine where one can expect that `naked' super-Earths to be found at orbits that are $< 1$ AU, has
to apply hydrodynamic mass loss calculations that do not neglect ionization, dissociation and recombination and consider all possible stellar EUV evolutionary tracks (Tu et al. 2015), as it was done for the habitable zone by Johnstone et al. (2015). This effort is beyond the scope of the present study but is planned to be carried out in the future.

\section{CONCLUSION}
We applied an 1D upper atmosphere EUV radiation absorption and hydrodynamic escape model that includes ionization, dissociation and recombination to hydrogen envelopes captured from protoplanetary nebulae surrounding rocky cores with masses between 1--5$M_{\oplus}$ at orbital locations of 0.1--1 AU. These different test planets have been exposed to a stellar EUV flux of a young solar-like star emitting 100 times more EUV radiation compared to present Sun. Depending on the assumed planetary parameters, the orbital distance, the corresponding EUV flux and the effective temperature, our model yields hydrogen escape rates of $\approx 10^{32}$ s$^{-1}$ to $10^{34}$ s$^{-1}$ and corresponding atmospheric mass loss rates of $\approx 10^{8}$ g s$^{-1}$ to $10^{10}$ g s$^{-1}$ between 1 AU and 0.1 AU, respectively. Our study also shows that the energy-limited formula can overestimate the atmospheric mass loss rates of hydrogen-dominated low mass planets such as `super-Earths' or sub-Neptunes especially at closer orbital distances up to a factor of $\approx 4$. For cooler planets with more compact atmospheres the difference between the hydrodynamic model and the loss rate obtained from eq. (1) is smaller. By assuming that $R_{\rm pl} \approx R_{0}\approx R_{\rm EUV}$ the energy-limited formula yields mass-loss rates too low by up to a factor of three in the studied parameter space.

\section*{ACKNOWLEDGMENTS}
The authors acknowledge the support by the FWF NFN project S11601-N16 `Pathways to Habitability: From Disks to Active Stars, Planets and Life', and the related FWF NFN subprojects, S11604-N16 'Radiation \& Wind Evolution from T Tauri Phase to ZAMS and Beyond', and S11607-N16 `Particle/Radiative Interactions with Upper Atmospheres of Planetary Bodies Under Extreme Stellar Conditions'. H. Lammer, P. Odert and N. V. Erkaev acknowledges also support from the FWF project P25256-N27 `Characterizing Stellar and Exoplanetary Environments via Modeling of Lyman-$\alpha$ Transit Observations of Hot Jupiters'. N. V. Erkaev acknowledges support by the RFBR grant No 15-05-00879-a. Finally, we thank an anonymous referee for his suggestions and recommendations that help to improve this work.

\appendix
\section{MODEL DESCRIPTION}
For studying the hydrogen loss of the test planets, we apply a
time-dependent 1-D hydrodynamic upper atmosphere model
that solves the system of the fluid equations for mass,
\begin{equation}
\frac{\partial \rho}{\partial t} + \frac{\partial({\rho v R^2})}{R^2\partial R} = 0,
\end{equation}
momentum,
\begin{eqnarray}
\frac{\partial \rho V }{\partial t} + \frac{\partial \left[ R^2(\rho V^2+P)\right]}{R^2\partial R} =
-\rho\frac{\partial U}{\partial R}  + 2 \frac{P}{R},
\end{eqnarray}
and energy conservation,
\begin{eqnarray}
\frac{\partial \left[\frac{1}{2}\rho V^2+ E +\rho U \right]}{\partial t}
+\frac{\partial V R^2\left[\frac{1}{2}\rho V^2 + E + P + \rho U \right] }{R^2\partial R} =\nonumber\\
 Q_{\rm EUV} - Q_{L_\alpha} + \frac{\partial }{R^2\partial R}\left(R^2 \chi \frac{\partial T}{\partial R}\right).
\end{eqnarray}
Here $Q_{\rm EUV}$ is the stellar EUV volume heating rate, which depends on the stellar EUV flux at
the orbital distance of the test planets and on the atmospheric density, and is given by
\begin{equation}
Q_{\rm EUV} = \eta \sigma_{\rm EUV}\left(n_{\rm H} + n_{\rm H_2}\right)\phi_{\rm EUV} ,
\end{equation}
$Q_{L_\alpha}$ is the Layman-alpha cooling, given by
\begin{equation}
Q_{L_\alpha} = 7.5\cdot 10^{-19} n_e n_{\rm H} \exp(-118348/T),
\end{equation}
 in units of (erg cm$^{-3}$ s$^{-1}$, $\chi$ is the thermal conductivity (Watson et al. ,1981)), given by
 \begin{equation}
 \chi = 4.45\cdot 10^4 \left(\frac{T}{1000}\right)^{0.7},
 \end{equation}
 $E$ is the thermal energy, given by
\begin{eqnarray}
E = \left[\frac{3}{2} (n_{\rm H} + n_{\rm H^+}) + \frac{5}{2}( n_{\rm H_2} + n_{\rm H_2^+})\right]kT.
\end{eqnarray}
and
U is the gravitational potential, given by
\begin{eqnarray}
U = \frac{G M_{\rm pl}}{R_{\rm 0}} \left( 1- \frac{R_{\rm 0}}{R}    \right).
\end{eqnarray}
Parameter $\eta$ is the ratio of the net local heating rate to the rate of the stellar radiative absorption. Generally, the value $\eta$ is not constant with altitude. Shematovich et al. (2014) studied the photolytic and electron impact processes in a hydrogen-dominated thermosphere by solving the kinetic
Boltzmann equation and by applying a Direct Simulation Monte Carlo model. From the calculated
energy deposition rates of the stellar EUV flux and that of the accompanying primary electrons
that are caused by electron impact processes in the H$_2 \rightarrow$H transition region in the upper
atmosphere, it was shown that $\eta$ varies between $\approx 10$\% and 20\% and does not reach
higher values than 20\% above the main thermosphere altitude, if photoelectron impact processes are included.

Because the current model does not self-consistently calculate $\eta$ with hight we assume
an average $\eta$ value of 15\%. This value is more realistic then those assumed by Penz et al. (2008)
of 60 \% or the 25 \% assumed by Jackson et al. (2012) and agrees also with the suggestion by Owen and Jackson (2012) that
estimates of total mass loss rates with $\eta\approx 30$\% are unrealistic high.

As in Murray-Clay et al. (2009), Erkaev
et al. (2013), Lammer et al. (2013) and Lammer et al. (2014), we assume a single wavelength for all photons ($h\nu = 20$ ev) and use an average
EUV photoabsorption
cross sections $\sigma_{\rm EUV}$ for hydrogen atoms and molecules about
$2\times10^{-18}$ cm$^{2}$  and $1.2\times10^{-18}$ cm$^{2}$, respectively.
The applied values are in agreement with experimental and theoretical data of Bates (1963),
Cook and Metzger (1964), and Beynon and Cairns (1965).

The continuity equations for the number densities of the atomic neutral hydrogen $n_{\rm H}$, atomic hydrogen ions $n_{\rm H^+}$,
hydrogen molecules $n_{\rm H_2}$, and hydrogen molecular ions $n_{\rm H_2^+}$, can then be written as
\begin{eqnarray}
\frac{\partial \left(n_{\rm H}\right)}{\partial t} + \frac{1}{R^2}\frac{\partial \left(n_{\rm H} v R^2\right)}{\partial R}=
 - \nu_{\rm H} n_{\rm H} - \nu_{\rm Hcol} n_e n_{\rm H} + \nonumber \\
 \alpha_{\rm H} n_{e}n_{\rm H^+} +
2\alpha_{\rm H_2}n_{\rm e}n_{\rm H_2^+}  + 2\nu_{dis} n_{\rm H_2}n - 2\gamma_{\rm H} n n_{\rm H}^2,
\end{eqnarray}
\begin{eqnarray}
\frac{\partial \left(n_{\rm H^+}\right)}{\partial t} + \frac{1}{R^2}\frac{\partial \left(n_{\rm H^+}v R^2\right)}{\partial R}=\nu_{\rm H} n_{\rm H} + \nu_{\rm Hcol} n_e n_{\rm H}- \nonumber \\
\alpha_{\rm H} n_{e}n_{\rm H^+},
\end{eqnarray}
\begin{eqnarray}
\frac{\partial \left(n_{\rm H_2}\right)}{\partial t} + \frac{1}{R^2}\frac{\partial \left(n_{\rm H_2}v R^2\right)}{\partial R}=
-\nu_{\rm H_2} n_{\rm H_2} - \nonumber \\
\nu_{dis} n_{\rm H_2}n +
\gamma_{\rm H} n n_{\rm H}^2,
\end{eqnarray}
\begin{eqnarray}
\frac{\partial \left(n_{\rm H_2^+}\right)}{\partial t} + \frac{1}{R^2}\frac{\partial \left(n_{\rm H_2^+}v R^2\right)}{\partial R}=\nu_{\rm H_2} n_{\rm H_2} - \alpha_{\rm H_2} n_{e}n_{\rm H_2^+} .
\end{eqnarray}
The electron density is determined for quasi-neutrality conditions
\begin{eqnarray}
n_{\rm e}=n_{H^+}+n_{H_2^+}
\end{eqnarray}
and the total hydrogen number density
\begin{eqnarray}
n=n_{\rm H} + n_{\rm H^+} + n_{\rm H_2} + n_{\rm H_2^+}.
\end{eqnarray}
$\alpha_{\rm H}$ is the recombination rate related to the reaction \mbox{H$^{+}+e\rightarrow H$} of $4\times 10^{-12} (300/T)^{0.64}$
cm$^{3}$ s$^{-1}$, $\alpha_{\rm H_2}$ is the dissociation rate of \mbox{H$_2^+ $+$e\rightarrow$H + H}: $\alpha_{\rm H_2}$=$2.3\times 10^{-8} (300/T)^{0.4}$
cm$^{3}$ s$^{-1}$,
$\nu_{diss}$ is the thermal dissociation rate of \mbox{H$_2$ $ \rightarrow$ H + H}: 1.5 $\cdot$ 10$^{-9}$ $\exp(-49000/T)$,
$\gamma_{\rm H}$ is the rate of reaction \mbox{H + H $\rightarrow$ H$_2$}: $\gamma_{\rm H}$ = 8.0 $\cdot $ 10$^{-33}$ (300/T)$^{0.6}$
(Yelle, 2004).

$\nu_{\rm H}$ is the
hydrogen ionization rate, and $\nu_{H_2}$ is the ionization rate of molecular hydrogen (Storey and Hummer 1995; Murray-Clay et al. 2009),
\begin{eqnarray}
\nu_{\rm H} = 5.9 \cdot 10^{-8} \phi_{\rm EUV} {\rm s^{-1}} , \quad \nu_{\rm H^2} = 3.3\cdot 10^{-8} \phi_{\rm EUV} {\rm s^{-1}},
\end{eqnarray}
and  $\nu_{\rm Hcol}$  is the collisional ionization rate (Black, 1981), $\nu_{\rm Hcol}$ = 5.9$\cdot 10^{-11} T^{1/2} \exp(-157809/T)$ .

$\phi_{\rm EUV}$ is the function describing the EUV flux absorption in the atmosphere
\begin{equation}
\phi_{\rm EUV}=\frac{1}{4\pi}\int_0^{\pi/2+\arccos(1/r)} J_{\rm EUV}(r,\theta)
2\pi\sin(\theta)d\theta.
\end{equation}
Here, $J_{\rm EUV}(r,\theta)$ is the function of spherical coordinates that describes the spatial variation of the  EUV flux  due to the atmospheric absorption (Erkaev et al. 2015), $r$ corresponds to the radial distance from the planetary
center noramalized to $R_{\rm 0}$ .

In the hydrodynamic equations, the mass density, $\rho$, and the pressure, $P$,
can then be written as
\begin{equation}
\rho=m_{\rm H}\left(n_{\rm H} + n_{\rm H^+}\right) + m_{\rm H_2}\left(n_{\rm H_2} + n_{\rm H_2^+}\right),
\end{equation}
\begin{equation}
P=\left(n_{\rm H}+n_{\rm H^+}+n_{\rm H_2}+n_{H_2^+}+n_{\rm e}\right)kT,
\end{equation}
where $T$ is the upper atmosphere temperature and $k$ is the  Boltzmann constant, and $m_{\rm H}$ and $m_{\rm H_2}$ are the
masses of the hydrogen atoms and molecules, respectively.

For atmospheres that are in long-term radiative equilibrium, the temperature $T_{\rm 0}$ near the lower boundary of the simulation domain is quite close to the planetary effective and equilibrium temperatures $T_{\rm eff}\approx T_{\rm eq}$. The hydrodynamic model is only applicable as long as enough collisions occur, which is the case if the Knudsen number is $< 0.1$. We set the upper boundary conditions in the supersonic region assuming the radial derivatives of the density, temperature and velocity are zero.

For computational convenience we introduce dimensionless quantities
\begin{eqnarray}
\tilde\rho=\rho /\rho_0, \quad \rho_0 = N_0 m_{\rm H_2} , \\
r = R/R_0, \quad   \tilde U = m_{\rm H} U/(k T_0), \\
\quad \tilde V= V/V_{T0},  \quad V_{T0}=\sqrt{k T_0/m_{\rm H}} ,  \\
\quad \tilde T= T/T_0, \tilde P = P/(\rho_0 V_{T0}^2),   \\
 \quad  X = m_{\rm H} n_{\rm H}/ \rho, \quad X^+ = m_{\rm H^+} n_{\rm H^+}/\rho, \\
\quad  Y = m_{\rm H_2} n_{\rm H_2}/\rho, \quad Y^+ = m_{\rm H_2^+} n_{\rm H_2^+}/\rho ,\\
  \tilde Q_{\rm EUV} = \eta \sigma_{EUV} \phi_{EUV}R_0 /(m_{\rm H_2}V_{T0}^3 ), \\
 \tilde Q_{L_\alpha} = 7.5\cdot 10^{-19} N_0 R_0 /(m_{\rm H_2} V_{T0}^3), \\
 \tilde \nu_{\rm H} =  \nu_{\rm H} R_0 /V_{T0} , \quad   \tilde \nu_{\rm H^2} =  \nu_{\rm H^2} R_0 /V_{T0},\\
 \tilde \alpha_{\rm H} =  \alpha_{\rm H} N_0 R_0 /V_{T0}, \quad
  \quad  \tilde \alpha_{\rm H^2} =  \alpha_{\rm H^2} N_0 R_0 /V_{T0},\\
  \tilde \nu_{\rm Hcol} =\nu_{\rm Hcol} N_0 R_0/V_{T0}, \quad
  \tilde \nu_{\rm diss} = \nu_{\rm diss}  N_0 R_0 /V_{T0}, \\
  \tilde \gamma_{\rm H} = \gamma_{\rm H} N_0^2 R_0 / V_{T0},
  \quad \tilde \chi = \chi T_0 / (\rho_0 V_{T0}^3 R_0).
   \end{eqnarray}
Subscript ``0'' denotes lower boundary values.
The normalized equations can be written as follows
\begin{eqnarray}
\frac{\partial \tilde \rho }{\partial t} + \frac{\partial \left( r^2\tilde \rho \tilde V\right)}{r^2\partial r} =0, \label{rho}\\
\frac{\partial \tilde \rho X }{\partial t} + \frac{\partial \left( r^2\tilde \rho X\tilde V\right)}{r^2\partial r} = \nonumber\\
-\tilde \nu_{\rm H} X \tilde \rho - \tilde \nu_{\rm Hcol} \tilde T^{1/2} \tilde \rho^2 X (2 X^+ + Y^+) + \nonumber\\
\tilde \alpha_{\rm H} \tilde\rho^2 X^+ (2 X^+ +Y^+) +\nonumber\\
\tilde \alpha_{\rm H^2} \tilde \rho^2 Y^+ (2 X^+ + Y^+) + \nu_{diss} \tilde \rho^2 Y (1+ X + X^+) \nonumber\\
-  \tilde \gamma_{\rm HH}  \tilde \rho^3 (1+X + X^+) X^2 .\\
\frac{\partial \tilde \rho X^+}{\partial t} + \frac{\partial \left( r^2\tilde \rho X^+\tilde V\right)}{r^2\partial r} = \nonumber\\
\tilde \nu_{\rm H} X \tilde \rho + \tilde \nu_{\rm Hcol} \tilde T^{1/2} \tilde \rho^2 X (2 X^+ + Y^+)\nonumber \\
-\tilde \alpha_{\rm H} \tilde\rho^2 X^+ (2 X^+ +Y^+), \\
\frac{\partial \tilde \rho Y^+ }{\partial t} + \frac{\partial \left( r^2\tilde \rho Y^+ \tilde V\right)}{r^2\partial r} = \nonumber\\
\tilde \nu_{\rm H^2} \tilde \rho Y  - \tilde \alpha_{\rm H^2} \tilde \rho^2 Y^+ (2 X^++ Y^+), \nonumber\\
\frac{\partial \tilde \rho \tilde V }{\partial t} + \frac{\partial \left[ r^2(\tilde \rho \tilde V^2+ \tilde P)\right]}{r^2\partial r} =\nonumber\\
-\tilde \rho\frac{\partial \tilde U}{\partial r}  + 2 \frac{\tilde P}{r}, \\
\frac{\partial \left[\frac{1}{2}\rho \tilde V^2+ \tilde E +
\tilde \rho \tilde U \right]}{\partial t} \nonumber\\
+\frac{\partial \tilde V r^2\left[\frac{1}{2}\rho \tilde V^2 + \tilde E + \tilde P + \tilde \rho \tilde U \right] }{r^2\partial r} =\nonumber\\
 \tilde Q_{\rm EUV} -\tilde Q_{L_\alpha} + \frac{\partial }{r^2\partial r}\left(r^2 \tilde \chi \frac{\partial \tilde T}{\partial r}\right), \\
 \tilde P = \tilde \rho \tilde T (1 +X + 3 X^+ + Y^+)/2.0, \\
  \tilde E = \tilde \rho \tilde V^2/2 +
 \tilde \rho \tilde T (5 + X + 7 X^+ + 3 Y^+)/4.0 . \label{E}
\end{eqnarray}

The obtained equations (\ref{rho}-\ref{E}) make a self-consistent closed system
with respect to 6 unknown quantities $ \rho, X, X^+, Y^+, T, V  $.
The seventh quantity $Y$ (ratio of the molecular hydrogen mass to the total mass) is determined by
simple equation
\begin{equation}
Y = 1 - X -X^+ -Y^+ .
\end{equation}
We apply the finite difference numerical scheme
of MacCormack to integrate the system of equations in time, which
can be written in a vector form
\begin{eqnarray}
\frac{\partial {U}}{\partial t} + \frac{\partial \Gamma(U)}{\partial r} = \Psi(U)
\end{eqnarray}
Finite difference approximation of this equation is the following
\begin{eqnarray}
 \bar{U}_i^{n+1} = U_i^n - \frac{\Delta t}{\Delta r}[ \Gamma(U_{i+1}^n) - \nonumber\\
 \Gamma(U_i^n) ]+ \Delta t \Psi(U_i^n), \\
U_i^{n+1} = \frac{1}{2}(\bar{U}_i^{n+1}  + U_i^n  ) - \frac{\Delta t}{2\Delta r}[\Gamma(\bar{U}_i^{n+1}) - \nonumber \\
\Gamma(\bar{U}_{i-1}^{n+1}) ]
 +\frac{\Delta t}{2} \Psi(\bar{U}_i^{n+1}).
\end{eqnarray}
This two-step MacCormack method is of second order approximation. It is
suitable for smooth solutions without sharp discontinuities.
A stationary radial distributions of the atmospheric quantities can be obtained
as a result of time relaxation.

\begin{figure}
\begin{center}
\includegraphics[width=0.95\columnwidth]{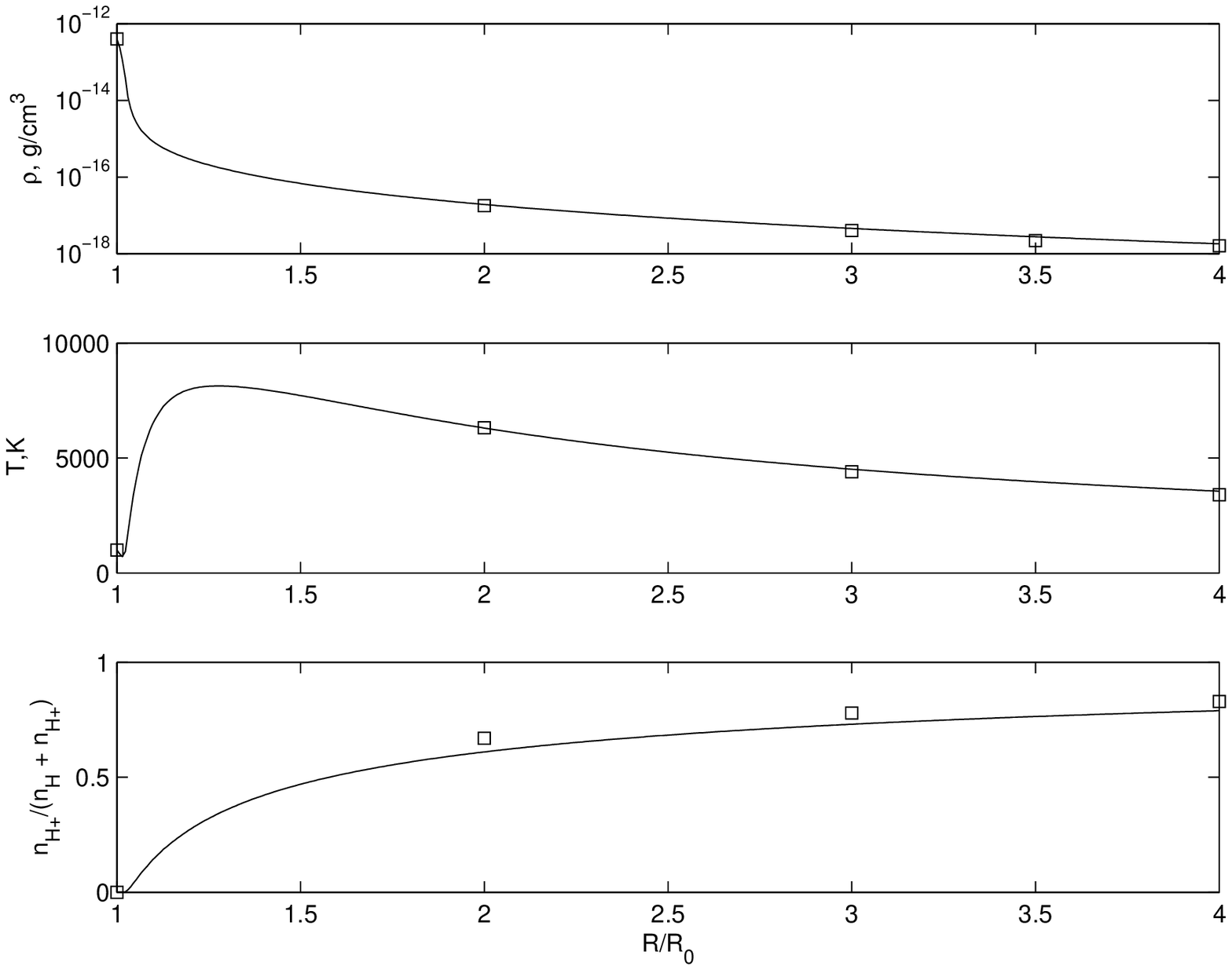}
\caption{Comparison between the results of atmospheric mass density, temperature and n$_{\rm H^+}/(n_{\rm H} + n_{\rm H^+}$ fraction as function of
distance for the hot Jupiter HD 209458b  of the applied hydrodynamic model (solid lines) to the solutions (square symbols) obtained by Murray-Clay et al. (2009)}
\end{center}
\end{figure}

At the lower boundary ($ r = 1$), we set the conditions to be
\begin{eqnarray}
\tilde T = 1, \quad \tilde \rho =1, \quad X =0, \quad X^+ =0, \quad Y^+=0.
\end{eqnarray}
The upper boundary is chosen in the supersonic outflow region, where we set zero
conditions for the radial derivatives of the quantities (i.e. free boundary conditions).

For tests of our numerical code, Erkaev et al. (2013) compared the calculated velocity profile
of an EUV-heated hydrogen-dominated upper atmosphere around an Earth-mass planet with the
analytical solution of Parker (1958) and obtained a perfect agreement (see Fig. 3 in Erkaev et al. 2013).
For validating our model for the application of the test planets in this study,
we compare the results obtained for the hot Jupiter HD 209458b with the results obtained for the same planet
of Murray-Clay et al. (2009). Murry-Clay et al. (2009) used an approach
that differed significantly from ours. These authors solved the boundary value problem for stationary hydrodynamic
equations with upper boundary conditions at the sonic point. They searched for a unique solution that
passes through the sonic point from the subsonic region to the supersonic one.

In our model we use the input parameters corresponding to the stellar and planetary
parameters of HD 209458b. A comparison of the atmospheric mass density, temperature and n$_{\rm H}^+/(n_{\rm H} + n_{\rm H^+}$ fraction
is shown in Figure A1. The solutions of our model results for HD 209458b correspond to the solid curves and
the solutions of Murray-Clay et al. (2009) correspond to the squares.
One can see that the comparison is rather good. The total escape rate obtained by Murray-Clay et al. (2009)
is $3.3 \cdot 10^{10}$ g s$^{-1}$. In our model the hydrogen loss rate of $3.33\cdot 10^{10}$ g s$^{-1}$ is nearly identical.
This very good agreement between the solutions obtained by different methods is a
convincing test to validate our numerical approach.
\end{document}